\documentstyle[preprint,prl,aps]{revtex}
\begin{document}
\preprint{TAUP 2385-96}
\draft
\title{Thermodynamics and the measure of entanglement}
\author{Sandu Popescu\thanks{E-mail:  \tt sp230@newton.cam.ac.uk}}
\address{Isaac Newton Institute, Cambridge University, 20 Clarkson Road,
Cambridge CB3 0EH, U.K.}
\author{Daniel Rohrlich\thanks{E-mail:  \tt daniel@ccsg.tau.ac.il}}
\address{School of Physics and Astronomy,
Tel Aviv University, Ramat Aviv 69978 Tel-Aviv, Israel}
\date{\today}
\maketitle
\begin{abstract}
We point out formal correspondences between thermodynamics and
entanglement.  By applying them to previous work, we show that {\it
entropy of entanglement} is the unique measure of entanglement
for pure states.
\end{abstract}
     Quantum entanglement is a most remarkable effect, and we have yet
to develop all the tools we need for studying it.  One tool we require
is a {\it measure} of entanglement\cite{shimony}.  How do we quantify
entanglement?  There have been many {\it  ad hoc} proposals for measuring
entanglement, but until now, the choice among them has been arbitrary.
Among the tools in use for studying entanglement are information
theory\cite{info} and various definitions of entropy\cite{BBPS,fb,BDSW}.  The
use of information theory and entropy to analyze entanglement indicates a
close correspondence between entanglement and thermodynamics.  In part,
the correspondence consists of common definitions, such as von Neumann
entropy.  In part, the correspondence is formal:  a formal principle of
thermodynamics may apply, {\it mutatis mutandis}, to the study of
entanglement.  The purpose of this note is to exploit the formal
correspondence:  we show that principles of thermodynamics, applied to
previous work on entanglement\cite{BBPS}, single out a unique measure of
entanglement for pure states.

     When Einstein searched for a universal formal principle from
which to derive a new mechanics (namely, special relativity) he took
for inspiration a general principle of thermodynamics:  The laws
of nature are such that it is impossible to construct a {\it perpetuum
mobile}\cite{einstein}.  This general principle (the second law) enabled
Carnot to show that all reversible heat engines operating between given
temperatures $T_1$ and $T_2$ are equally efficient.  Consider two
reversible heat engines; suppose that both absorb heat $Q_1$ at $T_1$
and expel heat $Q_2$ at $T_2$, but one does work $W$, and the other does
work $W^\prime >W$, per cycle.  The first engine, if run in reverse,
is a refrigerator---{\it absorbs} heat $Q_2$ at $T_2$ and {\it expels}
heat $Q_1$ at $T_1$---and requires only work $W$ per cycle.  Thus
the two engines together could provide $W^\prime -W$ in work per cycle
without changing their environment.  Such a conclusion contradicts the
second law, so both engines must do the same work:  $W=W^\prime$.

     The formal correspondence with entanglement is as follows:  The laws
of nature are such that it is impossible to create (or increase) entanglement
between remote quantum systems by local operations\cite{BBPS,fb}.  It is
clear that quantum mechanics does not allow local operations to create such
entanglement, although they may preserve or destroy entanglement\cite{note1}.
This general principle is the analogue of the second law of thermodynamics.
The analogue of a reversible heat engine is any reversible transformation,
consisting only of local operations, that transforms one entangled state
into another.  Let two experimenters, Alice and Bob, share pairs of quantum
systems in an entangled state.  One quantum system in each pair goes to
Alice, and the other goes to Bob.  In addition, each experimenter may have
access to other quantum systems that are not initially entangled.  Local
operation include any measurements or unitary transformations that Alice
performs on her systems, and that Bob performs on his.  Alice and Bob can
even exchange messages in the usual way (classical messages), but Alice
may not send Bob any system entangled with one that she keeps, nor may Bob
send such a system to Alice; thus, they cannot create entanglement between
their systems\cite{note2}.  In principle, however, this restriction still
allows them to transform entanglement in one pair of systems into
entanglement involving other systems.  For example, they may, using local
operations, transfer entanglement between two spins in a singlet state to
two identical spins that initially were uncorrelated, {\it i.e.} in a
product state, and the original spins will now be in a product state.  This
transformation is reversible:  local operations can transfer the singlet
entanglement back to the original spins, leaving the substitute spins
uncorrelated.

     We may consider more general transformations.  Suppose that Alice and
Bob share $k$ pairs of systems in an entangled state, and that, by local
operations only, they transform the entanglement to $n$ pairs of systems
in a different entangled state.  Since Bob and Alice have access to other
systems that are not initially entangled, $k$ and $n$ may be different.
Even if $n>k$, there need be no contradiction with the general principle
that it is impossible to create entanglement by local operations, because
the state of the $n$ pairs may be less entangled than the state of the
original $k$ pairs.  If Alice and Bob can transform $k$ pairs in one
entangled state into $n$ pairs in another entangled state {\it without
destroying any entanglement}, then any measure of entanglement must
assign the same entanglement to the $k$ initial pairs and the $n$ final
pairs.  But did they not destroy any entanglement?  That is, a question
arises with regard to the efficiency of the transformation:  could Alice
and Bob apply a different set of local operations to obtain the same number
$n$ of final pairs from a smaller number $k^\prime <k$ of initial pairs?

     The answer is that they cannot, {\it if both transformations are
reversible}.  For if it were possible to transform $k^\prime$ of the
initial pairs into $n$ of the final pairs by a different transformation,
Alice and Bob could then reverse the first transformation and transform
the $n$ pairs in the final state to $k$ pairs in the initial entangled
state.  In doing so, they would have added $k -k^\prime$ entangled pairs
to their initial supply, contradicting the general principle that it is
impossible to create entanglement by local operations\cite{note1}.  Thus
$k^\prime =k$.

     The reversible local transformations we have assumed are, in fact,
consistent with quantum mechanics. Bennett, Bernstein, Popescu and
Schumacher\cite{BBPS} have shown that is possible, with local operations
only, to transform $k$ systems in an entangled state $\vert\Psi^\prime_{AB}
\rangle$ into $n$ systems in a different entangled state $\vert \Psi_{AB}
\rangle$.  The transformation is reversible when the number of systems
becomes arbitrarily large.  That is, the ratio $n/k$ tends to a constant
in the limit $k\rightarrow \infty$.  We can then assign, to $k$ systems
in a pure entangled state $\vert \Psi_{AB} \rangle$, the same measure of
entanglement as (say) $n$ singlet pairs.  Thus the problem of defining a
measure of entanglement for $k$ pure states reduces to the problem of
defining a measure of entanglement for $n$ singlets.  At first, it might
seem that many such measures, such as $n$, $n^2$ and $e^n$, would be
admissible.  But actually, the measure must be proportional to $n$.  The
reason is that the transformations under consideration are reversible only
when the number of systems becomes arbitrarily large\cite{note3}.  Indeed,
the ratio $n/k$ nearly always tends to an irrational number\cite{BBPS}, and
if the number is irrational, we can never reversibly transform $n$ singlets
into a finite number $k$ of systems in the state $\vert \Psi_{AB} \rangle$.
Reversibility requires us to go to the limit of infinite $n$, and for
infinite $n$ there is no way to define total entanglement.  We can only
define entanglement {\it per system}.  Here too, thermodynamics provides
the formal principle:  the thermodynamic limit requires us to define
intensive quantities.  Likewise, the measure of entanglement must be
intensive, {\it i.e.} the measure of entanglement of $n$ singlets must
be proportional to $n$.  It immediately follows that the measure of
entanglement for pure states is unique (up to a constant factor).  Since
the measure of entanglement of $k$ systems $\vert \Psi_{AB} \rangle$
approaches the measure of entanglement of $n$ pairs in a singlet state
$\vert S_{AB} \rangle$, and since the measure is intensive, we have
$kE(\vert \Psi_{AB} \rangle ) =nE( \vert S_{AB} \rangle )$, where $E$
denotes the measure.  Thus
\begin{equation}
E(\vert \Psi_{AB} \rangle ) =\lim_{n,k \rightarrow \infty}
{n\over k} E( \vert S_{AB} \rangle )~~~,
\label{1}
\end{equation}
{\it i.e.} the entanglement of the state $\vert \Psi_{AB} \rangle $
is proportional to the number of singlet pairs per system in the state
$\vert \Psi_{AB} \rangle$, in this limit.  The proportionality constant
$E(\vert S_{AB}) \rangle )$---measuring the entanglement of a singlet
pair---simply defines a conventional unit, and we set it to 1.

     We have shown that the measure of entanglement for pure states
$\vert \Psi_{AB} \rangle$ is the limit $E(\vert \Psi_{AB} \rangle )$
of Eq.\ (\ref{1}).  It remains to compute this limit.  Indeed, Bennett,
Bernstein, Popescu and Schumacher\cite{BBPS} have computed it: $E(\vert
\Psi_{AB} \rangle )$ equals the {\it entropy of entanglement} of the
state $\vert\Psi_{AB} \rangle$.  The entropy of entanglement is the von
Neumann entropy of the partial density matrix seen by either Alice or
Bob, and equals the Shannon entropy of the squares of the coefficients of
the entangled state in the Schmidt decomposition\cite{BBPS}.  The entropy
of entanglement is zero for a pair of systems in a product state, and 1
for a pair of spin-1/2 particles in a singlet state; it is never negative.
The entropy of entanglement is intensive, as required:  if the measure of
entanglement of one pair in a state $\vert \Psi_{AB}\rangle$ is $E (\vert
\Psi_{AB}\rangle )$, then the measure of entanglement of $k$ pairs in
the same state is $k\cdot E (\vert\Psi_{AB} \rangle )$.  Bennett,
Bernstein, Popescu and Schumacher\cite{BBPS} argue that the entropy of
entanglement is a good measure of entanglement for pure states, because
local operations can interconvert states of equal entropy of entanglement
with asymptotically perfect efficiency, but can never increase the entropy
of entanglement.  We have presented a stronger argument, showing that
it is the {\it unique} measure of entanglement for pure states.

     We arrive at this unique measure by considering all possible local
operations on entangled systems.  But, in a given setting, the possible
local operations may not all be practical.  For example, Alice and Bob
may not have access to arbitrarily many entangled systems.  Then various
measures of entanglement that are not valid in principle may be valid
(and useful) in practice, simply because Alice and Bob cannot perform the
operations that invalidate them.  These measures must not increase under
the local operations Alice and Bob can perform.  A formal statement of
this constraint is that if Alice and Bob can, by local operations,
transform an entangled state $\vert \Psi_{AB}\rangle$ into entangled
states $\vert \Phi^{(i)}_{AB} \rangle$ with probabilities $p_i$, respectively,
then any measure $E$ must satisfy $E (\vert \Psi_{AB} \rangle ) \ge \sum_i
p_i E(\vert \Phi^{(i)}_{AB} \rangle )$.  This constraint on measures of
entanglement is necessary, but not sufficient when all local transformations
are allowed.

     We have treated pure states.  What about density matrices?  Is
there a unique measure of entanglement for them?  Here, too, we can
apply thermodynamic principles.  Clearly, if there is a reversible
transformation between any density matrix and any reference state (such
as a singlet), our derivation of the measure of entanglement for pure
states extends immediately to density matrices.  At present, such
reversible transformations are an open question.  Two measures of
entanglement are particularly natural:  the entropy of formation and
the entropy of distillation\cite{fb}.  The former measures the number
of singlets Alice and Bob need in order to create a mixed state, while
the latter measures the number of singlets that Alice and Bob can extract
from a mixed state.  If there is a reversible transformation between
the entropy of formation and the entropy of distillation, then these
measures coincide (as they do for pure states) and define the unique
measure of entanglement for mixed states.

\acknowledgments
We thank A. Shimony for comments.  D. R. acknowledges support from the
State of Israel, Ministry of Immigrant Absorption, Center for Absorption
in Science (Giladi Program).

\section*{Note Added}
After completing this work, we learned of a paper by V. Vedral, M. B.
Plenio, M.A. Rippin and P. L. Knight\cite{vprk} which assumes, as we do,
that local operations cannot create entanglement.  Their paper presents
a whole class of `good' entanglement measures.  The reason that Vedral
{\it et al.} do not obtain a unique measure of entanglement, as we do,
is that they consider only those local operations that Alice and Bob may
perform on each of their entangled pairs in isolation from the other
pairs.  By contrast, here we consider collective local operations on the
pairs, and these imply much stronger constraints on the measure.


\begin{references}
\bibitem{note2}
By convention, Alice and Bob may exchange classical messages.  The logic
of this convention is as follows.  Local operations cannot increase
entanglement, but they also cannot increase classical correlations.  Thus,
if we define entanglement as whatever local operations cannot increase,
our definition does not distinguish between entanglement and classical
correlations.
\bibitem{note3}
We note that quantum mechanics does not allow local operations to increase
entanglement even in the limit $n,k\rightarrow\infty$.  Suppose initially
there are $M$ copies of an entangled state $\vert \Psi \rangle$, and finally
there are $M$ copies of the same state $\vert \Psi \rangle$ plus $M$ copies
of a new entangled state $\vert \Psi^\prime \rangle$.  The scalar product
between the initial and final states has norm less than 1, and in the limit
$M \rightarrow\infty$ it vanishes, unless $\vert \Psi^\prime \rangle$
reduces to a product state.   Thus, even in the limit of infinitely many
systems, there is no way to produce a final state with more entanglement
per system than the initial state.
\end{references}
\end{document}